\documentclass{egpubl}
\usepackage{eurovis2024}

\EuroVisShort  % for EuroVis additional material
\CGFccby
\usepackage[T1]{fontenc}
\usepackage{dfadobe}  

\usepackage{cite}  % comment out for biblatex with backend=biber
\BibtexOrBiblatex
\electronicVersion
\PrintedOrElectronic
\ifpdf \usepackage[pdftex]{graphicx} \pdfcompresslevel=9
\else \usepackage[dvips]{graphicx} \fi

\usepackage{egweblnk}
\usepackage{xcolor}
\usepackage{enumitem}
\usepackage{fontawesome}
\usepackage{colortbl}
\usepackage{graphicx}
\usepackage{array}
\usepackage{float}
\usepackage{pdflscape}

\usepackage{xcolor}
\usepackage{enumitem}
\usepackage{fontawesome}
\usepackage{colortbl}
\usepackage{graphicx}
\usepackage{array}
\usepackage{float}

\newcolumntype{R}[1]{>{\raggedright\arraybackslash}p{#1}}
\newcommand{\pheading}[1]{\noindent\textbf{#1}}

\arrayrulecolor{gray}

\usepackage{tabularx}
\usepackage{tabu}                      % only used for the table example
\newcolumntype{P}[1]{>{\arraybackslash}p{#1}}

\usepackage{xurl}

\title[Mixing Modes]%
      {Mixing Modes: Active and Passive Integration of Speech, Text, and Visualization for Communicating Data Uncertainty}
\author[C. Stokes, C. Sanker, B. Cogley, \& V. Setlur]
{\parbox{\textwidth}{\centering Chase Stokes$^{1,2}$\orcid{0000-0001-7644-9021},
        Chelsea Sanker$^{3}$\orcid{0000-0001-6106-9587},
        Bridget Cogley$^{4}$\orcid{0009-0002-0125-5171},
        and Vidya Setlur$^{1}$\orcid{0000-0003-3722-406X}
        }
        \\
{\parbox{\textwidth}{\centering 
    $^1$Tableau Research, Palo Alto, USA\\
    $^2$University of California, Berkeley, USA\\
    $^3$Stanford University, Palo Alto, USA\\
    $^4$Versalytix, USA
       }
}
}

\begin{document}

% uncomment for using teaser

% \teaser{
% \includegraphics[width=\textwidth]{figures/teaser.pdf}
%   \caption{
 
%   }
%   \label{fig:teaser}
% }

\maketitle

%-------------------------------------------------------------------------

\begin{abstract}
Interpreting uncertain data can be difficult, particularly if the data presentation is complex. We investigate the efficacy of different modalities for representing data and how to combine the strengths of each modality to facilitate the communication of data uncertainty. We implemented two multimodal prototypes to explore the design space of integrating speech, text, and visualization elements. A preliminary evaluation with 20 participants from academic and industry communities demonstrates that there exists no one-size-fits-all approach for uncertainty communication strategies; rather, the effectiveness of conveying uncertain data is intertwined with user preferences and situational context, necessitating a more refined, multimodal strategy for future interface design. Materials for this paper can be found on \textcolor{blue}{\href{https://osf.io/6g8ex/}{OSF}}.
   
%-------------------------------------------------------------------------

%  ACM CCS 1998
%  (see https://www.acm.org/publications/computing-classification-system/1998)
% \begin{classification} % according to https://www.acm.org/publications/computing-classification-system/1998
% \CCScat{Computer Graphics}{I.3.3}{Picture/Image Generation}{Line and curve generation}
% \end{classification}

%-------------------------------------------------------------------------

%  ACM CCS 2012
%   (see https://www.acm.org/publications/class-2012)
%The tool at \url{http://dl.acm.org/ccs.cfm} can be used to generate
% CCS codes.


\end{abstract}  

%-------------------------------------------------------------------------

\section{Introduction}
\label{section:intro}
Communicating data uncertainty can be challenging, especially for audiences with limited statistical expertise~\cite{MacEachren:2005, hullman2019authors}. 
However, accurately communicating uncertainty is critical for decision-making, as it could impact risk assessments. 

Prior work has explored visual representations for uncertain data \cite{kay2016ish, fernandes2018uncertainty, padilla2021uncertain}. Other modes of communication, such as text or speech explanations, can mitigate some of these issues but have limitations. For instance, audiences with limited graphical literacy more easily interpret text, but text cannot convey statistical information. ~\cite{schriver:1997}. Speech conveys tone not present in written text, but is ephemeral and poses challenges for conveying complex information~\cite{Sperber1995-SPER}. Integrating visual, textual, and speech elements may use one method's strengths to offset another's limitations, facilitating more effective data communication.

Depending on the user's context, different modes may be used for decision-making. For example, text-to-speech features can be useful when the user's hands are occupied, such as while driving. However, these passive contexts limit users' ability to further interact with the data. We implemented two multimodal interface types: passive, which provided an integrated presentation of information while minimizing user effort, and active, which emphasized user-driven interaction
% , enabling modification 
and in-depth exploration~\cite{endert:2012}. We then completed a preliminary evaluation of the prototypes, focusing on design implications and future research directions. 
% This design probe helps to inform tools and media for communicating data uncertainty to a target audience.

\section{Related Work}
\label{section:related_work}

\begin{figure*}[ht]
    \centering
    \includegraphics[width = 0.9\linewidth]{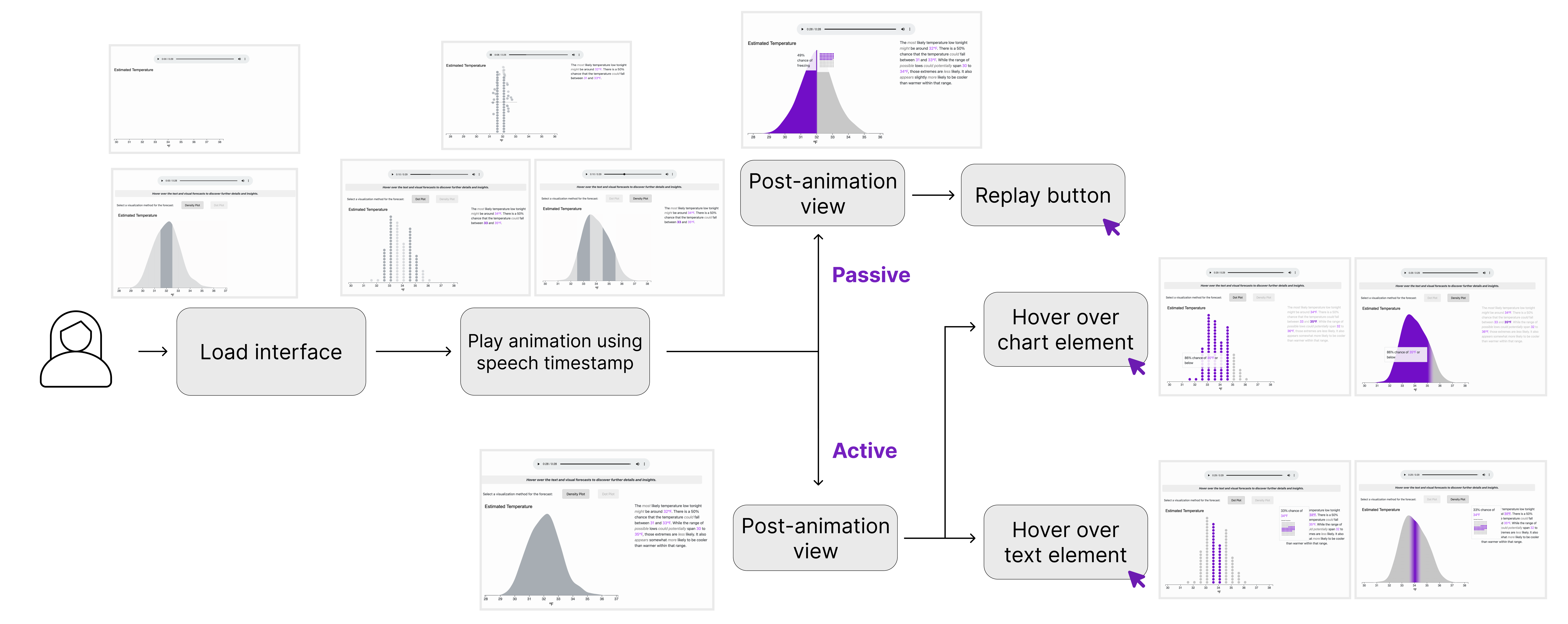}
    \caption{Diagram showing multimodal components and possible user interactions with passive and active interfaces. 
    }
    \label{fig:prototype_diagram}
\end{figure*}

\subsection{Representations of Uncertainty}
A range of studies have examined the use of uncertainty visualizations, from surveys ~\cite{Bonneau2014OverviewAS, padillabook:2021, skeels:2008} to specialized tools~\cite{jena:2020} and libraries~\cite{ggdist}. This research includes applications such as weather forecasting \cite{padilla2021uncertain} and public transportation \cite{fernandes2018uncertainty, kay2016ish}. Different kinds of uncertainty visualizations have been evaluated for decision-making \cite{padilla2021uncertain, fernandes2018uncertainty, kay2016ish}. 

Text representations of uncertainty often use hedges to signal uncertainty, such as `might' or `possibly'~\cite{lakoff:1973, SzarvasEtAl2012}. 
These hedges can reflect either inherent uncertainty in the data being described or the author's uncertainty about it. Text descriptions can also convey statistics about uncertain outcomes.

Hedges are also used in speech along with phonetic patterns like speech rate, pauses, and pitch \cite{SchererEtAl1973, JiangPell2015}. Uncertain speech often has rising intonation \cite{SmithClark1993}, slower speech rate \cite{SchererEtAl1973, SmithClark1993}, and more frequent pauses  \cite{SchererEtAl1973}. 
% The relationship between pitch and uncertainty is less consistent; 
Some studies find a higher pitch associated with lower confidence \cite{JiangPell2015}, while others find a lower pitch associated with lower confidence \cite{SchererEtAl1973}.  
These acoustic characteristics can also impact how confident or uncertain a listener deems the speaker \cite{kirkland2022s}. 

Building upon this rich body of research, our work further explores the integration of speech, text, and visualization through the implementation of multimodal prototypes for both passive and active consumption of uncertainty communication.

\subsection{Multimodal Techniques for Communicating Information}
Existing research has explored how multimodal interaction can blend various modes of communication, such as visual, auditory, and haptic cues, to effectively convey information. Bromley and Setlur~\cite{bromley2023difference} explored the use of language with visualizations by employing hedges to describe trends (e.g., ``\textit{gradual} fall''). 

Multimodal techniques can be useful for conveying uncertainty to users who may be blind or have low vision. Sharif et al.~\cite{sharif2023conveying} provide empirical findings from semi-structured interviews with 16 screenreader users to identify user preferences in uncertainty visualizations. Several studies explore how non-speech audio can be used to represent data, either on its own or in combination with visualizations~\cite{listen:1996,LodhaEtAl1997,Bearman2011}. Previous work has also explored differences between passive and active communication of data insights. Both variants result in similar levels of comprehension \cite{Rbbelen2021InteractiveVS}; interactivity can increase user enjoyment \cite{dyer2008animated}, while static visualizations offer a more direct message \cite{mahajan2018comparative}.

\section{Prototype Implementation}
\label{section:system}

Our work builds upon this research to explore the research question, \textbf{``What are user preferences for multimodal presentations of uncertain data?''} 
We implemented two prototypes with passive and active interface behaviors. Design choices were made in consultation with a UX designer at a visual analytics company with 10+ years of experience and one of the co-authors who is an interactive visualization designer with 15+ years of experience. Prototype design was iterative, with feedback provided on four occasions.

\subsection{Decision-making Context}
The interfaces studied here were situated in a decision-making context. 
Users assumed the role of a road maintenance company contracted to treat roads with salt to prevent icing. Users were tasked with applying salt to the roads when the temperature was at or below 32ºF (0ºC) to prevent ice from forming. They were informed that salt supplies were limited, so maintenance companies must balance cost (and supply) with damage prevention. To make this decision, they viewed and listened to a forecast depicting predicted temperature lows for a given evening. 

% Whether a user made the ``right'' decision depends on the costs of salting and possible damages, which were not available in this scenario. 
The appropriateness of a user's decision to salt the roads depends on the costs of salting and potential damages; information not provided in this scenario. The decision context was used to provide a \textit{situation} of use for the interface rather than to assess rationality. Prior work examined each mode of information in decision-making, finding that speech led to higher trust in information, while text was associated with lower decision confidence \cite{stokes2024delays}. This related paper studied uncertainty data in the same decision context as our work, but differed in research questions and methodology.

\begin{figure*}[ht]
    \centering
    \includegraphics[width = 0.85\linewidth]{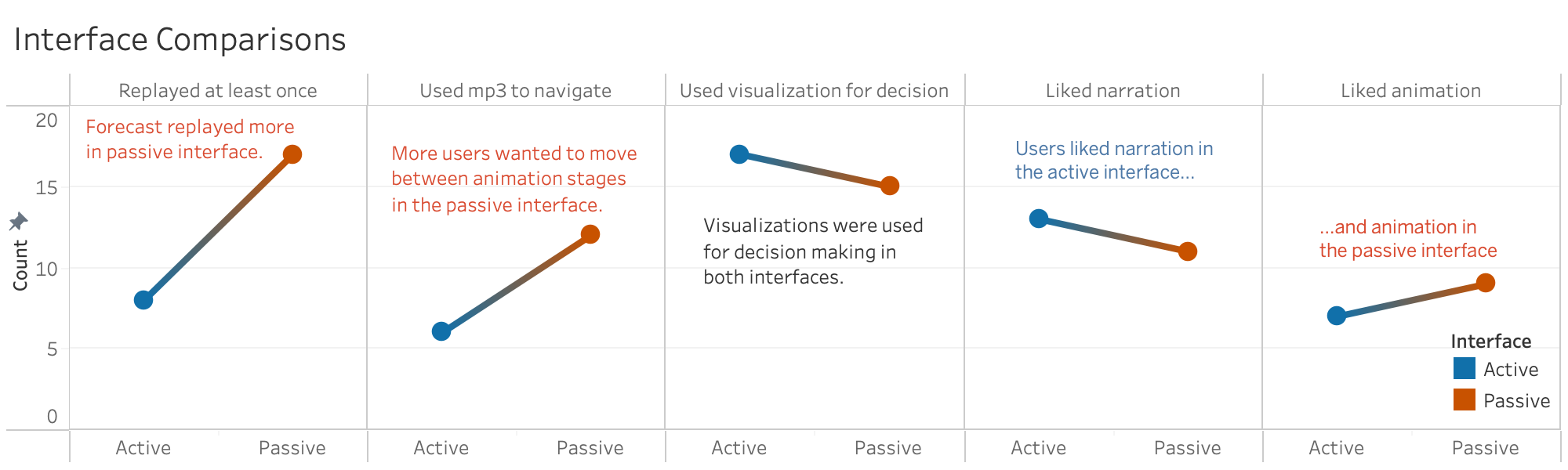}
    \caption{
    Insights from comparing active and passive interfaces. Different features may be better suited for specific interfaces, while some user interactions are consistent. 
    }
    \label{fig:interface_comparison}
\end{figure*}

\subsection{System Components}

The interface flow for both prototypes is shown in Figure \ref{fig:prototype_diagram}. Each prototype employs visualizations, text descriptions with hedges, and speech elements with pauses and rate changes to convey data uncertainty. The speech forecast is displayed at the top of the page. A D3 visualization~\cite{2011-d3} and a text description of uncertainty are shown below the MP3 player. An event listener in the interface monitors time progression in the speech module, triggering animated updates in the visualization and text modules. In addition, the active prototype leverages interaction with the visual and textual modes, including additional linking between the two.

\subsubsection{Data Module}
Raw distribution data is integrated into a Node.js~\cite{Node} application using an Express framework, serving the dataset as a static file. Upon loading the prototype, a random trial is selected from the dataset. This selection process links the columns of the .csv file to the unique trial number. Dataset values in the .csv are derived by selecting 100 points from a normal distribution, such that the resulting distributions may not have been normally distributed. This variation provided more ecologically valid distributions, as natural data tends to not always follow a normal distribution. 

Text templates and timings also serve as a static JavaScript file, including each sentence and the timestamp it occurs in the speech element. The text-to-speech synthesis engine computes the timing information while generating speech from the SSML syntax. These timings are available as static files. 

\subsubsection{Multimodal Modules}
\pheading{Visualization module.}
To communicate data uncertainty in visualizations, we employ density and 100-quantile dot plots \cite{kay2016ish}. Density plots are created using a kernel density estimator to determine the continuous probability curve from discrete data points. The quantile dot plot is created using a histogram, with 20 evenly spaced bins based on the range of the x-axis scale. %For each bin, the bin height is computed, and the requisite number of dot icons is added using SVG circle elements.

\pheading{Text module.}
The text descriptions are generated with natural language templates containing hedges and summary statistics from the visualized dataset \cite{stokes2024delays}. These statistics include the mean of the distribution, the range of the middle 50\% of data, the full range of the data, and a verbal representation of distribution skew. The skew value is computed using the skewness function in R \cite{moments}, then mapped to magnitudes (``slightly'' to ``significantly'') and a positive or negative direction. Hedges were included in each sentence to communicate uncertainty (e.g., `might,' `could'). A standard black color is used to render the text, with a gray color (\#757575) applied to hedges to indicate a higher level of uncertainty. Colors employed in the prototypes are compliant with the WCAG AA guidelines for color contrast~\cite{w3c_wai_2019}.

\pheading{Speech module.}
The text templates are translated into Google Speech Synthesis Markup Language (SSML)~\cite{googlessml} to provide adjustments in pitch, rate of speech, and pauses for communicating uncertainty. For hedges in the text, we modify the speech to $65\%$ of the original rate and lower the pitch by $5\%$. We apply the same pitch treatment to numerical values, slowing the speech to $70\%$ and adding a $0.2$ second break before the values. 

\subsubsection{Interaction Module}
In both prototypes, the forecast can be replayed to view and listen to the information. During the replay, the animated sequence for the text and visualization also repeats. The active prototype includes detail-on-demand tooltips and linking between modes, shown in Figure \ref{fig:prototype_diagram}. 
There are three distinct hover interactions (hedges, numerical values, visualization) that provide dynamic feedback and detailed insights about the data. A video demonstrating these behaviors is included in the supplementary material.

When a user hovers over a hedge in the text description, the word visually responds by ``wobbling'' at $3$º angles and applying a $0.5$px blur effect
Hovering over a numerical value in the text shows a tooltip featuring an icon array visualizing the likelihood of the hovered number occurring within the dataset. Simultaneously, the corresponding section of the visualization that represents this number is highlighted and wobbles to draw attention and convey uncertainty. Upon hovering over a mark in the visualization, a tooltip appears with a text description of the cumulative likelihood of achieving the corresponding value or lower.

\section{Preliminary User Study}
\label{section:user_study}
Using the passive and active prototypes as a design probe, we conducted a preliminary evaluation to qualitatively assess the overarching idea of combining speech, text, and visualization modes for communicating data uncertainty information. 

\subsection{Method}
We employed a within-subjects design to elicit subjective user feedback for both interfaces. Participants engaged with both prototypes. Half of the participants viewed the active prototype first; half viewed the passive. Sessions lasted approximately 45 minutes and were recorded with the participant's consent. One author coded the transcripts in close consultation with a second author.

For each interface, participants viewed the speech-linked animation, accompanied by the narration and linked text transcript. They could replay the animation and narration or interact with available features. They were encouraged to think aloud to share their observations and reactions to the interface. The facilitator then asked for specifics about the potential use of the information and possible changes to the interface. After participants had viewed both interfaces, the facilitator asked about their usage of the different modes and their perception of each interface's strengths and weaknesses. 

\subsection{Participants}

The $20$ participants ($8$ female, $12$ male) were fluent in English. Half were graduate students from a nearby university, and half were professionals employed at a technology enterprise with a variety of backgrounds (e.g., technical engineering, product management). Participants were recruited via internal mailing lists and Slack channels. Graduate students were compensated with a \$20 Amazon gift card. Professionals were not compensated due to company policy prohibiting payment. We refer to participants using [$P$\#].

\subsection{Study Findings}
Code counts from the interview transcripts can be found in Figure \ref{fig:interface_comparison}, along with supplementary and recruitment materials. Overall, participants used the speech and text modes as supplementary to the visualization for both interfaces. Interacting with the active interface was helpful for decision-making and providing further detail on event likelihoods. We observed a slight difference in interaction behaviors: passive-first participants replayed the speech element more ($1.5$ times on average) than active-first ($0.6$ times).

\pheading{Replay and navigation.} Replay frequency and interaction with the speech element were much higher for the passive interface ($17$ participants) than the active interface ($8$ participants). 
When viewing the passive interface, $12$ participants used the speech element to navigate between animation stages, compared to $6$ for the active interface. Some of the participants felt that there were too many elements in the interface ($13$ active, $11$ passive).

\pheading{Trends in interaction.}
In the active interface, participants could choose the visualization displayed (dot plot or density plot), switching an average of $11.5$ times. 
$14$ participants preferred the density plot, citing familiarity with the chart type. Each participant interacted with the density plot an average of $6.9$ times, taking an average of $14.5$ seconds (s) per interaction. Similarly, they interacted with the dot plot an average of $6.6$ times with an average of $10.2$s per interaction. 
The text component had fewer interactions, an average of $5.2$ interactions per participant, but each interaction was longer, averaging $30.0$s.  

% Interactions with these visualizations are summarized in Table \ref{tab:interactions}.

% \begin{table}
%     \centering
%     \begin{tabular}{|l|l|l|}
%     \hline
%         \textbf{Feature} & \textbf{Avg. Frequency} & \textbf{Average Duration (s) }\\
%         \hline
%         Density Plot & 6.9 times per participant & 14.5 s\\
%         \hline
%         Dot Plot & 6.6 times per participant &  10.2 s\\
%         \hline
%         Text Passage & 5.2 times per participant &  30.0 s\\
%         \hline
%     \end{tabular}
%     \caption{Summary data for participant interactions}
%     \label{tab:interactions}
% \end{table}

\pheading{Decision-making.}
Participants frequently reported that they used visualizations in the decision-making process ($17$ active, $15$ passive). 
The interactions with the visualizations were particularly useful, with $13$ participants reporting that they relied primarily on information from the visualization tooltips. According to $P5$, ``Some of the interactions, if not all of them, I feel, were more valuable... I got to the decision much faster.'' 

The text tooltips were less useful overall, with $9$ participants finding them redundant and not decision-relevant.
Rather than offering the likelihood of going \textit{below} a given temperature, as in the visualization tooltips, the text tooltips offered the likelihood of being \textit{at} a given temperature.
$9$ participants found this relatively unhelpful, as this information did not greatly inform or change their decisions. However, $9$ participants still found the interaction beneficial for overall understanding - ``the information from decision-making perspective doesn't seem super helpful... but I absolutely love seeing the vis.'' [$P10$].

\section{Discussion and Future Work}
\label{section:discusssion}
The exploration of these two interfaces provided useful insights for future multimodal system design. The active interface provided users with a rich set of interactions, and participants often switched between visualizations, suggesting that providing more than one way to represent the data visually was useful. However, for both interfaces, the narrative and animation straddled the line between distracting and engaging. These varied responses underscore the challenges in designing clear multimodal data representations.

\pheading{Benefits of each mode.} 
Participants most often used visualizations in the decision-making process. Many of the participants also liked the narration as a source of information - ``it's very useful if you're too lazy to read the text... Audio complemented with text is very powerful'' [$P14$]. Speech also improved comprehension of the data display; ``the audio was helpful to understand the overall instructions of how to read the view'' [$P20$]. 
These findings are consistent with prior results in decision-making with unimodal data representations for visualization and speech: \textbf{decision quality was highest with visualization; trust was highest with speech} \cite{stokes2024delays}.

The staged chart composition illustrated how to read the chart, which was useful for some users: ``The animation is quite cool, and I understand how to read the graph if I don't know how'' [$P14$]. However, several participants admitted that the multimodality was overwhelming. $P1$ commented, ``I couldn't keep track of all of that information... there was just too much going on.'' Future work should investigate how to balance multiple modes but not overwhelm users with too many interface components.  \textbf{Different speakers, text descriptions, and visualizations could also affect user experiences and preferences.}

\pheading{Context of use.} 
Participants noted crucial differences between the potential usage context of the interfaces. The passive interface was viewed as better suited for a more casual setting - `` I think if it was not a serious decision... I would actually prefer the [passive]... I could just see it and then just draw insight from it quickly'' [$P16$], or may also be better suited for users with less experience in data; $P6$ suggested, ``That would be something that would probably appeal to a lot of folks who don't have a desire to go in and be analysts themselves..''
The active interface seemed a better fit for an audience with some expertise in similar types of data interpretation, particularly those who are ``proficient in the tool already'' [$P12$].

Future work should explore multimodality in different settings and with users of different levels of expertise. \textbf{While casual users might appreciate a guided, narrative experience with high-level insights (the passive interface), expert users may seek detailed, interactive tools for data exploration (the active interface).}

\section{Conclusion}
This work discusses two interfaces that explore the paradigms of expressing data uncertainty in both active and passive consumption contexts. Preliminary user feedback illuminated use cases and challenges when presenting and interpreting multimodal data uncertainty across these contexts.
This research identifies the need for refined, context-specific multimodal strategies in uncertainty communication, different situational contexts (e.g., for analysis vs. during travel), and user groups (e.g., data experts vs. casual users).

\newpage
\clearpage

% bibtex

\bibliographystyle{eg-alpha-doi}

\bibliography{0_bib}

\newcommand{\etalchar}[1]{$^{#1}$}
\begin{thebibliography}{\uppercase{KKHM16}}

\bibitem[Bea11]{Bearman2011}
\textsc{Bearman N.}:
\newblock Using {S}ound to {R}epresent {U}ncertainty in {F}uture {C}limate {P}redictions for the {UK}.
\newblock In \emph{17th International Conference on Auditory Display} (2011), International Community for Auditory Display.

\bibitem[BHJ{\etalchar{*}}14]{Bonneau2014OverviewAS}
\textsc{Bonneau G.-P., Hege H.-C., Johnson C.~R., de~Oliveira~Neto M.~M., Potter K.~C., Rheingans P., Schultz T.}:
\newblock Overview and {S}tate-of-the-{A}rt of {U}ncertainty {V}isualization.
\newblock In \emph{Scientific Visualization: Uncertainty, Multifield, Biomedical, and Scalable Visualization} (London, 2014), Hansen C.~D., Chen M., Johnson C.~R., Kaufman A.~E., Hagen H., (Eds.), Springer, pp.~3--27.

\bibitem[BOH11]{2011-d3}
\textsc{Bostock M., Ogievetsky V., Heer J.}:
\newblock {D3: Data-Driven Documents}.
\newblock \emph{IEEE Transactions on Visualization and Computer Graphics 17}, 12 (2011), 2301--2309.

\bibitem[BS23]{bromley2023difference}
\textsc{Bromley D., Setlur V.}:
\newblock What {I}s the {D}ifference {B}etween a {M}ountain and a {M}olehill? {Q}uantifying {S}emantic {L}abeling of {V}isual {F}eatures in {L}ine {C}harts.
\newblock In \emph{IEEE Transactions on Visualization and Computer Graphics} (2023).

\bibitem[DAV08]{dyer2008animated}
\textsc{Dyer S., Adamo-Villani N.}:
\newblock {Animated Versus Static User Interfaces: A Study of Mathsigner™}.
\newblock \emph{International Journal of Human and Social Sciences 3}, 6 (2008).

\bibitem[EFN12]{endert:2012}
\textsc{Endert A., Fiaux P., North C.}:
\newblock Semantic {I}nteraction for {V}isual {T}ext {A}nalytics.
\newblock In \emph{Proceedings of the SIGCHI Conference on Human Factors in Computing Systems} (New York, NY, USA, 2012), CHI '12, Association for Computing Machinery, p.~473–482.

\bibitem[FWM{\etalchar{*}}18]{fernandes2018uncertainty}
\textsc{Fernandes M., Walls L., Munson S., Hullman J., Kay M.}:
\newblock Uncertainty {D}isplays {U}sing {Q}uantile {D}otplots or {CDF}s {I}mprove {T}ransit {D}ecision-{M}aking.
\newblock In \emph{Proceedings of the 2018 CHI Conference on Human Factors in Computing Systems} (New York, NY, USA, 2018), Association for Computing Machinery, pp.~1--12.

\bibitem[Goo24]{googlessml}
\textsc{Google}:
\newblock Speech {S}ynthesis {M}arkup {L}anguage ({SSML}) {R}eference, 2024.
\newblock Google Cloud Documentation.
\newblock URL: \url{https://cloud.google.com/text-to-speech/docs/ssml}.

\bibitem[Hul19]{hullman2019authors}
\textsc{Hullman J.}:
\newblock Why authors don't visualize uncertainty.
\newblock \emph{IEEE transactions on visualization and computer graphics 26}, 1 (2019), 130--139.

\bibitem[JED{\etalchar{*}}20]{jena:2020}
\textsc{Jena A., Engelke U., Dwyer T., Raiamanickam V., Paris C.}:
\newblock Uncertainty {V}isualisation: {A}n {I}nteractive {V}isual {S}urvey.
\newblock In \emph{2020 IEEE Pacific Visualization Symposium (PacificVis)} (2020), IEEE, pp.~201--205.

\bibitem[JP15]{JiangPell2015}
\textsc{Jiang X., Pell M.~D.}:
\newblock On {H}ow the {B}rain {D}ecodes {V}ocal {C}ues {A}bout {S}peaker {C}onfidence.
\newblock \emph{Cortex 66} (2015), 9--34.

\bibitem[Kay23]{ggdist}
\textsc{Kay M.}:
\newblock \emph{{ggdist}: {V}isualizations of {D}istributions and {U}ncertainty}.
\newblock Northwestern University, 2023.
\newblock R package version 3.3.0.
\newblock URL: \url{https://mjskay.github.io/ggdist/}.

\bibitem[KKHM16]{kay2016ish}
\textsc{Kay M., Kola T., Hullman J.~R., Munson S.~A.}:
\newblock When ({I}sh) is {M}y {B}us? {U}ser-{C}entered {V}isualizations of {U}ncertainty in {E}veryday, {M}obile {P}redictive {S}ystems.
\newblock In \emph{Proceedings of the 2016 CHI Conference on Human Factors in Computing Systems} (New York, NY, USA, 2016), Association for Computing Machinery, pp.~5092--5103.

\bibitem[KLSG22]{kirkland2022s}
\textsc{Kirkland A., Lameris H., Sz{\'e}kely E., Gustafson J.}:
\newblock Where’s the {U}h, {H}esitation? {T}he {I}nterplay {B}etween {F}illed {P}ause {L}ocation, {S}peech {R}ate and {F}undamental {F}requency in {P}erception of {C}onfidence.
\newblock In \emph{Proceedings of Interspeech} (Incheon, Korea, 2022), Interspeech, pp.~4990--4994.

\bibitem[KN22]{moments}
\textsc{Komsta L., Novomestky F.}:
\newblock \emph{Moments, {C}umulants, {S}kewness, {K}urtosis and {R}elated {T}ests}.
\newblock CRAN, 2022.
\newblock R package version 0.14.1.
\newblock URL: \url{https://cran.r-project.org/web/packages/moments/moments.pdf}.

\bibitem[Lak73]{lakoff:1973}
\textsc{Lakoff G.}:
\newblock Hedges: {A} {S}tudy in {M}eaning {C}riteria and the {L}ogic of {F}uzzy {C}oncepts.
\newblock \emph{Journal of Philosophical Logic 2}, 4 (1973), 458--508.

\bibitem[LBH{\etalchar{*}}97]{LodhaEtAl1997}
\textsc{Lodha S.~K., Beahan J., Heppe T., Joseph A., Zane-Ulman B.}:
\newblock {MUSE}: {A} {M}usical {D}ata {S}onification {T}oolkit.
\newblock In \emph{Proceedings of the 4th International Conference on Auditory Display, Palo Alto, California, November 2--5, 1997} (1997), Georgia Institute of Technology.

\bibitem[LWS96]{listen:1996}
\textsc{Lodha S., Wilson C., Sheehan R.}:
\newblock {LISTEN}: {S}ounding {U}ncertainty {V}isualization.
\newblock In \emph{Proceedings of Seventh Annual IEEE Visualization '96} (1996), IEEE, pp.~189--195.

\bibitem[MG18]{mahajan2018comparative}
\textsc{Mahajan K.~N., Gokhale L.~A.}:
\newblock {Comparative Study of Static and Interactive Visualization Approaches}.
\newblock \emph{International Journal on Computer Science and Engineering (IJCSE), e-ISSN} (2018), 0975--3397.

\bibitem[MRH{\etalchar{*}}05]{MacEachren:2005}
\textsc{MacEachren A., Robinson A., Hopper S., Gardner S., Murray R., Gahegan M., Hetzler E.}:
\newblock Visualizing {G}eospatial {I}nformation {U}ncertainty: {W}hat {W}e {K}now and {W}hat {W}e {N}eed to {K}now.
\newblock \emph{Cartography and Geographic Information Science 32} (07 2005), 139--160.

\bibitem[{Ope}23]{Node}
\textsc{{OpenJS Foundation}}:
\newblock Node.js, 2023.
\newblock \url{https://nodejs.org}.

\bibitem[PKH21]{padillabook:2021}
\textsc{Padilla L., Kay M., Hullman J.}:
\newblock Uncertainty {V}isualization.
\newblock In \emph{Wiley StatsRef: Statistics Reference Online}. Wiley, 02 2021, pp.~1--18.

\bibitem[PPKH21]{padilla2021uncertain}
\textsc{Padilla L.~M., Powell M., Kay M., Hullman J.}:
\newblock Uncertain about {U}ncertainty: {H}ow {Q}ualitative {E}xpressions of {F}orecaster {C}onfidence {I}mpact {D}ecision-making with {U}ncertainty {V}isualizations.
\newblock \emph{Frontiers in Psychology 11} (2021), 579267.

\bibitem[RSK{\etalchar{*}}21]{Rbbelen2021InteractiveVS}
\textsc{R{\"o}bbelen A., Schmieding M.~L., Kopka M., Balzer F., Feufel M.~A.}:
\newblock {Interactive Versus Static Decision Support Tools for COVID-19: Randomized Controlled Trial}.
\newblock \emph{JMIR Public Health and Surveillance 8} (2021).
\newblock URL: \url{https://api.semanticscholar.org/CorpusID:248240389}.

\bibitem[SC93]{SmithClark1993}
\textsc{Smith V.~L., Clark H.~H.}:
\newblock On the {C}ourse of {A}nswering {Q}uestions.
\newblock \emph{Journal of Memory and Language 32} (1993), 25--38.

\bibitem[Sch97]{schriver:1997}
\textsc{Schriver K.~A.}:
\newblock \emph{Dynamics in {D}ocument {D}esign: {C}reating {T}ext for {R}eaders}.
\newblock John Wiley \& Sons, Inc., USA, 1997.

\bibitem[SLSR08]{skeels:2008}
\textsc{Skeels M., Lee B., Smith G., Robertson G.}:
\newblock Revealing {U}ncertainty for {I}nformation {V}isualization.
\newblock In \emph{AVI '08: Proceedings of the working conference on advanced visual interfaces} (05 2008), vol.~9, pp.~376--379.

\bibitem[SLW73]{SchererEtAl1973}
\textsc{Scherer K.~R., London H., Wolf J.~J.}:
\newblock The {V}oice of {C}onfidence: {P}aralinguistic {C}ues and {A}udience {E}valuation.
\newblock \emph{Journal of Research in Personality 7} (1973), 31--44.

\bibitem[SSCS24]{stokes2024delays}
\textsc{Stokes C., Sanker C., Cogley B., Setlur V.}:
\newblock {From Delays to Densities: Exploring Data Uncertainty through Speech, Text, and Visualization}.
\newblock In \emph{Eurographics Conference on Visualization (EuroVis)} (2024), The Eurographics Association.
\newblock to appear.

\bibitem[SVF{\etalchar{*}}12]{SzarvasEtAl2012}
\textsc{Szarvas G., Vincze V., Farkas R., M{\'o}ra G., Gurevych I.}:
\newblock Cross-genre and {C}ross-domain {D}etection of {S}emantic {U}ncertainty.
\newblock \emph{Computational Linguistics 38}, 2 (2012), 335--367.

\bibitem[SW95]{Sperber1995-SPER}
\textsc{Sperber D., Wilson D.}:
\newblock \emph{Relevance: {C}ommunication and {C}ognition}.
\newblock Blackwell, Oxford, 1986/1995.

\bibitem[SZW23]{sharif2023conveying}
\textsc{Sharif A., Zhong R., Wang Y.}:
\newblock Conveying {U}ncertainty in {D}ata {V}isualizations to {S}creen-{R}eader {U}sers {T}hrough {N}on-{V}isual {M}eans.
\newblock In \emph{The 25th International ACM SIGACCESS Conference on Computers and Accessibility (ASSETS '23)} (New York, NY, USA, 2023), Association for Computing Machinery.

\bibitem[(W319]{w3c_wai_2019}
\textsc{(W3C) W. W. W.~C.}:
\newblock {WAI} {Web} {Accessibility} {Tutorials}: {Complex} {Images}.
\newblock World Wide Web Consortium website, 2019.
\newblock \url{https://www.w3.org/WAI/tutorials/images/complex/}.

\end{thebibliography}

%-------------------------------------------------------------------------

\newpage

\end{document}